\documentclass[journal]{IEEEtran}
\usepackage{amsmath}
\usepackage{graphicx}
\usepackage[bookmarks=false]{hyperref}

\begin{document}
\title{Effects of High Charge Densities \\ in Multi-GEM Detectors}
\author{
S.~Franchino$^{1}$,
D.~Gonzalez Diaz$^{1}$,
R.~Hall-Wilton$^{2}$,
H.~Muller$^{1}$,
E.~Oliveri$^{1}$,
D.~Pfeiffer$^{1,2}$,
F.~Resnati$^{*,1,2}$,
L.~Ropelewski$^{1}$,
M.~Van~Stenis$^{1}$,
C.~Streli$^{3}$,
P.~Thuiner$^{1,3}$,
R.~Veenhof$^{1}$\\
$^1$ CERN, Geneva, Switzerland.
$^2$ ESS, Lund, Sweden.
$^3$ TUW, Wien, Austria.
\thanks{$^*$ filippo.resnati@cern.ch}
}

\maketitle
\thispagestyle{empty}

\begin{abstract}
A comprehensive study, supported by systematic measurements and numerical computations, of the intrinsic limits of multi-GEM detectors when exposed to very high particle fluxes or operated at very large gains is presented.
The observed variations of the gain, of the ion back-flow, and of the pulse height spectra are explained in terms of the effects of the spatial distribution of positive ions and their movement throughout the amplification structure.
The intrinsic dynamic character of the processes involved imposes the use of a non-standard simulation tool for the interpretation of the measurements.
Computations done with a Finite Element Analysis software reproduce the observed behaviour of the detector.
The impact of this detailed description of the detector in extreme conditions is multiple:
it clarifies some detector behaviours already observed, it helps in defining intrinsic limits of the GEM technology, and it suggests ways to extend them.
\end{abstract}


\section{Introduction}
\IEEEPARstart{T}{he} wide spread of the Gaseous Electron Multiplier (GEM)~\cite{Sauli:1997} is due to several key features, chiefly the ability to withstand large particle fluxes at high and stable gains.
For this reason, GEMs are employed in several High Energy Physics experiments, and their use is foreseen for the upgrade of two of the large LHC experiments:
GEMs will be installed in the forward muon detector system of CMS~\cite{CMS}, and as the amplification device in the ALICE TPC~\cite{ALICE}.

This article is devoted to the detailed study of the intrinsic limitations of multi-stage GEM detectors, manifesting themselves in extreme conditions, far beyond the actual requirements for the upgrade of the LHC experiments.
A puzzling increase of the effective gain of a triple GEM detector at high fluxes was observed for the first time in 2006~\cite{Everaerts:2006}.
The extreme particle flux at which the effect occurred, however, deviated the attention towards more practical problems.
Recently, ALICE Collaboration measured a flux dependence of the ion back-flow in multi-GEM detectors~\cite{Ball:2014}.
Although both effects could be perceived as related to the same common origin, such a link could not be established.
In the following paragraphs we present the results of measurements at very high X-ray fluxes and very high gains, and we give an interpretation of the observed behaviours.
In section~\ref{sec:setup} the setup used is described.
In section~\ref{sec:measurement} we present the experimental observations, without details on the interpretation, that, instead, are given in section~\ref{sec:interpretation} together with the description of the model developed.

\section{Setup}
\label{sec:setup}
\begin{figure}[!t]
\centering
\includegraphics[width=3.5in]{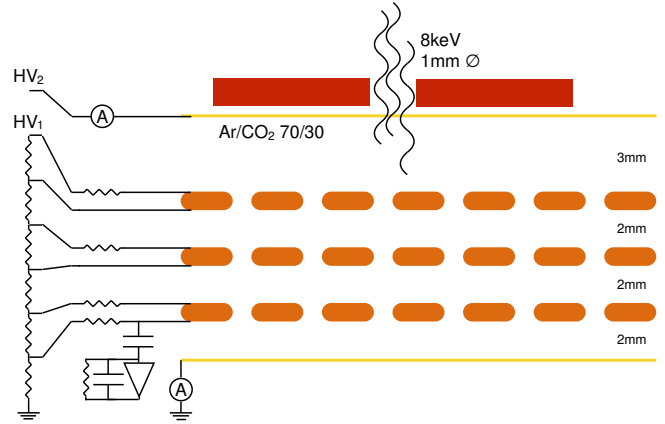}
\caption{Schematic representation of the setup.}
\label{fig:scheme}
\end{figure}

The detector used for the measurements is a \emph{standard} triple GEM~\cite{Altunbas:2002} flushed with Ar/CO$_2$~70/30 mixture at 5~stl/h.
A GEM foil is a thin insulator (50~$\mu$m thick polyimide foil) metal-clad on both faces and with holes (typically 70~$\mu$m in diameter at a pitch of 140~$\mu$m) chemically etched.
A \emph{standard} triple GEM consists of three GEM foils at a distance of 2~mm from each other (transfer regions), with a conversion volume 3~mm long on top of the first GEM delimited by the cathode, and an induction region 2~mm long below the last GEM delimited by the anode (see image in figure~\ref{fig:scheme}).
The potentials to the GEM electrodes are provided through a resistor divider chain.
The cathode is powered independently in order to measure the ionic current.
The anode is connected to ground via a pico-ammeter (Keithley~6487).
The event by event signals are picked up from the bottom electrode of the last GEM.
A decoupling capacitor of 2.2~nF connects the bottom electrode of the last GEM to a charge preamplifier (ORTEC 142) and a shaper (ORTEC 450) chain, finally connected to a Multi Channel Analyser (AMPTEK 8000D).
The detector is irradiated with a collimated beam (1~mm in diameter) of 8~keV photons from a copper X-ray generator impinging orthogonally to the GEMs.
The primary electrons created in the conversion volume undergo three stages of amplification before reaching the anode.
The rate of interactions, mostly happening in the conversion volume, can be tuned by modifying the intensity of the X-ray generator.
The detector is also exposed to an $^{55}$Fe source for the high gain measurements.

\section{Measurements}
\label{sec:measurement}
\subsection{High flux}
\begin{figure}[!t]
\centering
\includegraphics[width=3.5in]{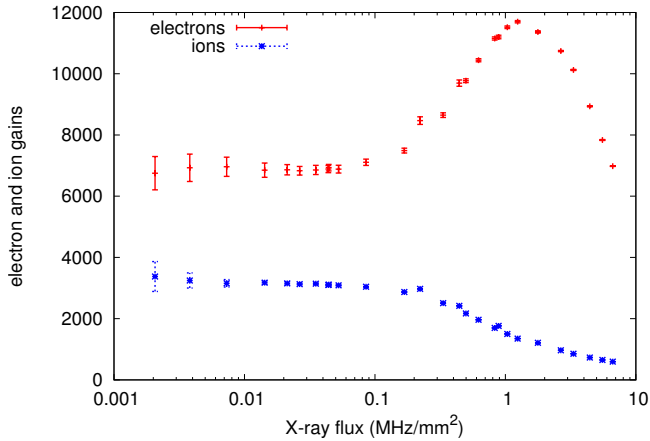}
\caption{Electron and ion average effective gains as a function of the detected X-ray flux. NB: 8~keV released in Ar/CO$_2$ 70/30 mixture produce about 290~electron-ion pairs, about ten times more than minimum ionising particles traversing 3~mm.}
\label{fig:gainVsFlux}
\end{figure}
The average effective gain of a triple GEM detector as a function of the detected X-ray flux shows first an increase followed by a decrease, as can be seen in figure~\ref{fig:gainVsFlux}.
This effect appears at larger fluxes the smaller the initial gain is, because the effect is related to the presence of ionic charges and the consequent electric field modification, as explained next.
As an illustrative example of a full set of systematic measurements, that will be published elsewhere, only the results from a particular setting will be described hereafter.
For this configuration, with a nominal effective gain of $7\times10^3$, the \emph{ion} gain is also shown in figure~\ref{fig:gainVsFlux}.
The electron (ion) gain is defined as the current at the anode (cathode) normalised to the primary ionisation charge and to the interaction rate.
With about 290~primary electrons per event, the electron gain, being stable up to 100~kHz/mm$^2$, exhibits a remarkable rise while, at the same time, the ion gain decreases.
The electron gain reaches a maximum at about 1~MHz/mm$^2$, after which it decreases together with the ion gain.
The ion back-flow (not shown), defined as the ratio of the cathode to anode currents, is stable up to 100~kHz/mm$^2$ at around 0.45, it decreases, and it stabilises again after 2~MHz/mm$^2$ at around 0.1.




\subsection{High gain}
\begin{figure}[!t]
\centering
\includegraphics[width=3.5in]{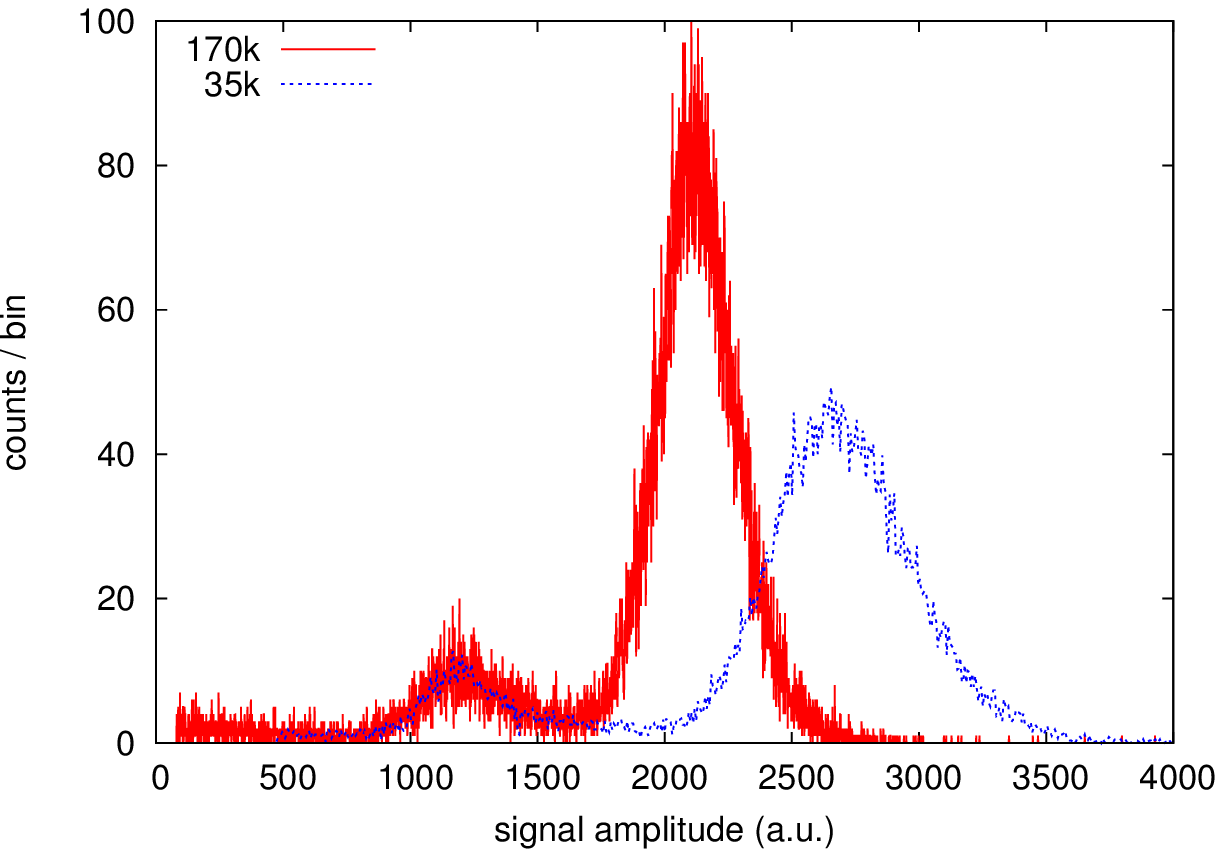}
\includegraphics[width=3.5in]{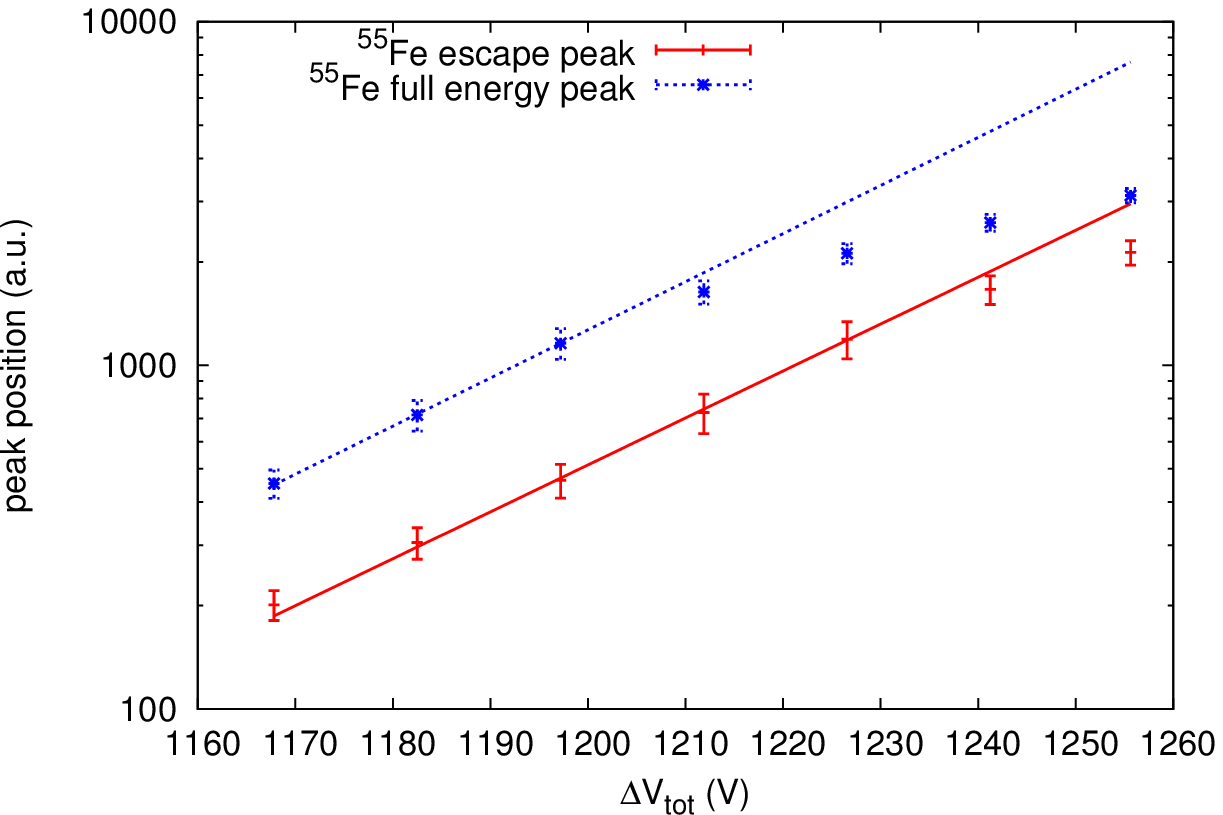}
\caption{Top: scaled spectra of the $^{55}$Fe source for moderately high gain of about $35\times10^3$ and very high gain of about $170\times10^3$. Bottom: position of the full energy peak and escape peak using an $^{55}$Fe source as a function of the total voltage across the three GEMs.}
\label{fig:gainVsVoltage}
\end{figure}
Figure~\ref{fig:gainVsVoltage} (bottom) shows the position of the full energy peak and escape peak using an $^{55}$Fe source (at a very low interaction flux) in the pulse height spectra varying the voltage across the GEMs.
The gain curve, expected to be exponential, shows a saturation for the 5.9~keV line starting at 1210~V, that corresponds to an effective gain larger than 10$^5$.
The saturation of the escape line appears at larger gains, suggesting that the phenomenon is linked to the charges within individual events.
A consistent saturation behaviour is observed in the average gain computed from the current at the anode.
The pulse hight spectra in figure~\ref{fig:gainVsVoltage} (top) are scaled such that the escape peaks appear in the same position.
We consider the spectrum at a moderate high gain of about $35\times10^3$ not affected by the saturation.
Instead, in the spectrum at very high gain of about $170\times10^3$ the saturation of the full energy peak is evident.
The saturation may affect already the escape peak, nevertheless what is shown in the figure is that the saturation is more important the larger the avalanche is.
The width of the saturated full energy peak is anomalously small, suggesting that even within very similar events the larger the avalanche is the more saturated it is.
The gain could not be pushed farther than what shown in figure~\ref{fig:gainVsVoltage} (bottom) because of the onset of discharges.

\section{Interpretation}
\label{sec:interpretation}
In the previous section we described two observations, not obviously related, but that can be both explained considering the variations of the electric field produced by the presence of positive ions and their movement.

The dynamic equilibrium of creation and evacuation of ions results in a steady space charge distribution in the transfer gaps and in the conversion volume that modify the electric field.
Considering a single GEM as a uniform ion generator located at $z = 0$, the electric field $E_z$ along $z$ (the unique non vanishing component of the electric field) in the transfer/conversion region of length $\delta$ is
\begin{align*}
E_z & = \sqrt{2J(z+z_0)/(\epsilon\mu)}, \text{with $z_0$ such that}\\
\Delta V & = \int_0^\delta E_z dz = \sqrt{8J/(9\epsilon\mu)}\left((\delta+z_0)^{3/2}-z_0^{3/2}\right),
\end{align*}
where $\Delta V$ is the voltage difference applied across the gap, $\mu$ is the ion mobility (the ion velocity is proportional to the electric field), $\epsilon$ is the permittivity of the gas,
and $J$ is the ion current density, that depends on the X-ray flux, the amount of primary ionisation, and the gain.
The modification results in a field that is no longer uniform: despite being always along the $z$ direction, its absolute value decreases where the positive ions \emph{enter} the volume and increases where the ions \emph{exit}.
The effect due to the electrons, drifting much faster than the ions, is completely negligible.
In reality, $J$ depends on $E_z$ because the electron collection in the holes and the ion extraction from the holes depend on this field; nevertheless, by measuring $J$, $E_z$ can be evaluated.
In a triple GEM stack, this effect translates into a change of the transfer and drift fields: as a function of the X-ray flux, the field at the \emph{entrance}\footnote{Entrance and exit are referred to the paths of the electrons.} of the GEMs decreases, and the one at the \emph{exit} increases.
This asymmetry results in both a more effective extraction of the electrons from one GEM and collection into the next one, i.e.\ the electron transmission through the GEMs, being less than 100\%~\cite{Bachmann:1999}, increases.
On the contrary but for the same reasons, the ion transmission decreases.
These considerations suffice to describe the increase (decrease) of the effective electron (ion) gain as a function of the X-ray flux.
\begin{figure}[!t]
\centering
\includegraphics[width=3.5in]{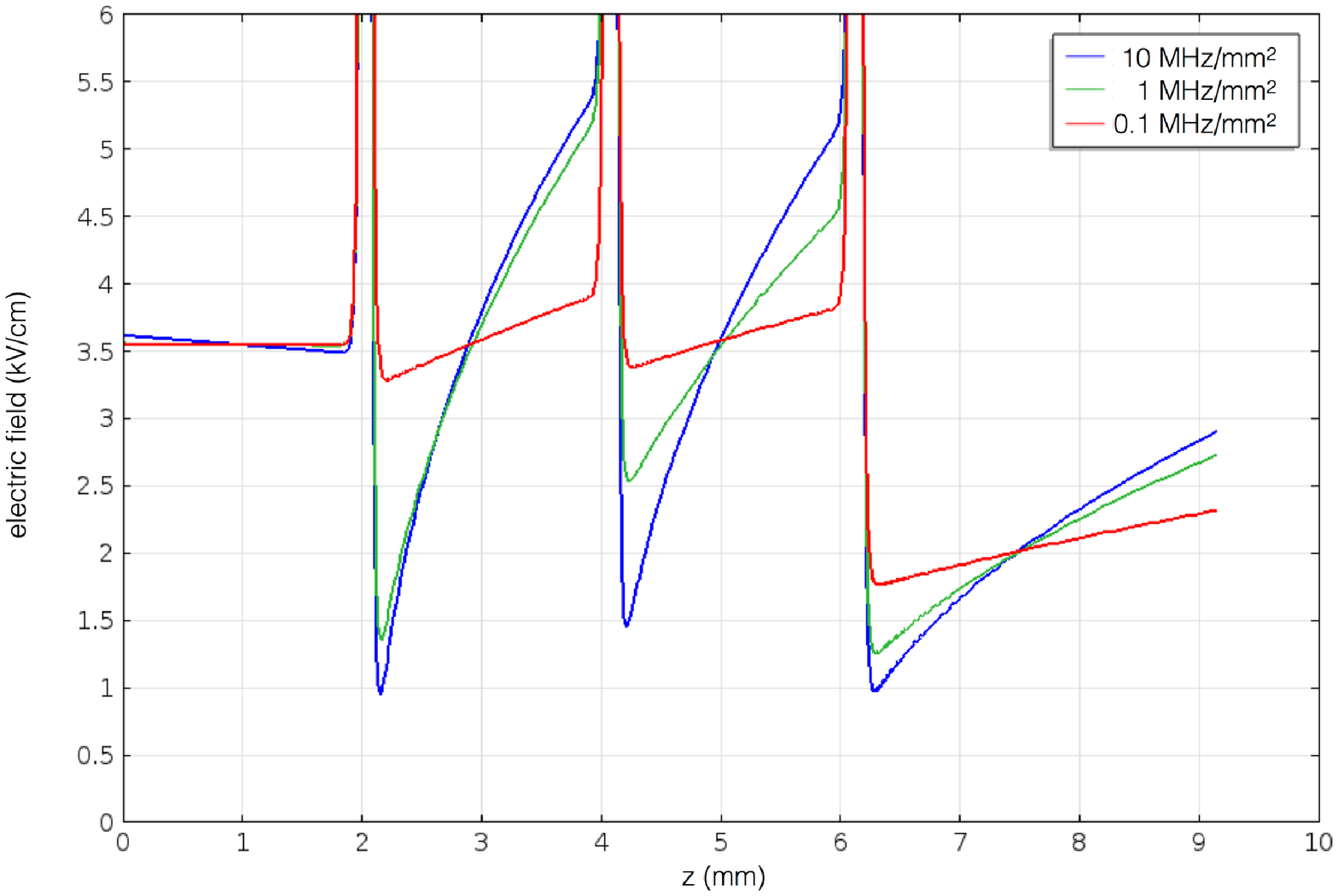}
\includegraphics[width=3.5in]{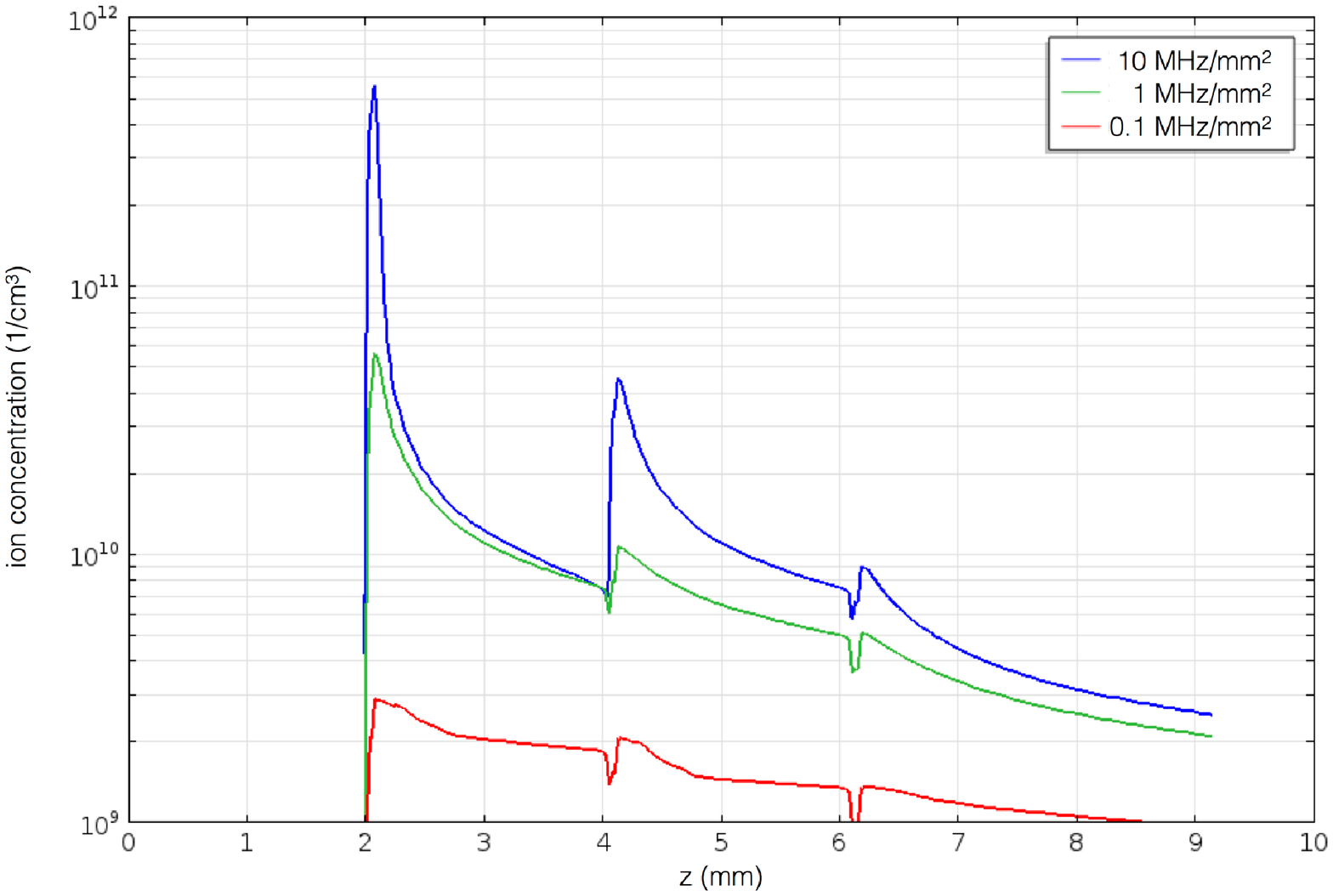}
\caption{The computed distortion of the electric field (top) and the steady state ion distribution (bottom) along the centre of the GEM holes for different interacting X-ray fluxes. The anode is at coordinate z = 0~mm.}
\label{fig:distortion}
\end{figure}

For complex geometries and in order to take into account the circular dependence of $J$ and $E_z$, the problem can be solved using a Finite Element Analysis program, such as COMSOL Multiphysics~\cite{COMSOL}, which allows to dynamically compute the electric field in the presence of charges, as well as the amplification and transport of the charges themselves under the action of the electric field.
COMSOL numerically approximates the solution of a set of second order partial differential equations, that in our case is composed of the drift and diffusion equations for electron and ion distributions, and the Poisson equation for the potential distribution.
The equations describe the electron and ion densities \emph{macroscopically},
therefore the stochastic behaviour of the avalanche cannot be taken into account.
The electron transport coefficients in Ar/CO$_2$~70/30 are computed using Magboltz~8.97~\cite{Biagi:1999}, the Townsend coefficient corrected for the Penning transfer is taken from~\cite{Sahin:2014}, and the ion mobilities are taken from~\cite{Ellis:1976}.
In order to retain the computation time, the 3D geometry is approximated with a 2D axisymmetric geometry where a single GEM hole is implemented.
The three GEMs have therefore the holes aligned.
The electric field and the ion density along the hole axis are shown in figure~\ref{fig:distortion} for different X-ray fluxes.
The transfer and the drift field are distorted, as described previously, yielding an increase of the effective gain.
An increase of the effective gain is expected in any system where the efficiency of transferring the electrons from one stage to the next is not 100\%.
In a double GEM this effect is present, but it is milder.
In a single GEM, where the electron collection efficiency from the conversion volume to the holes can reach 100\%, this effect is absent.
The measurement was performed up to 2~MHz/mm$^2$ at a gain of about $10^3$.
The variation of the drift field in this case, though present, has no impact on the effective gain, but it affects the ion gain which slightly decreases.

The reduction of the electric field at the \emph{entrance} of the GEM implies a lower speed of the ions and a consequent accumulation of positive charge, as shown in figure~\ref{fig:distortion} (bottom).
This results in a reduction of the amplification field in the last GEM hole that for large X-ray fluxes becomes sizeable and affects the multiplication process.
\begin{figure}[!t]
\centering
\includegraphics[width=3.5in]{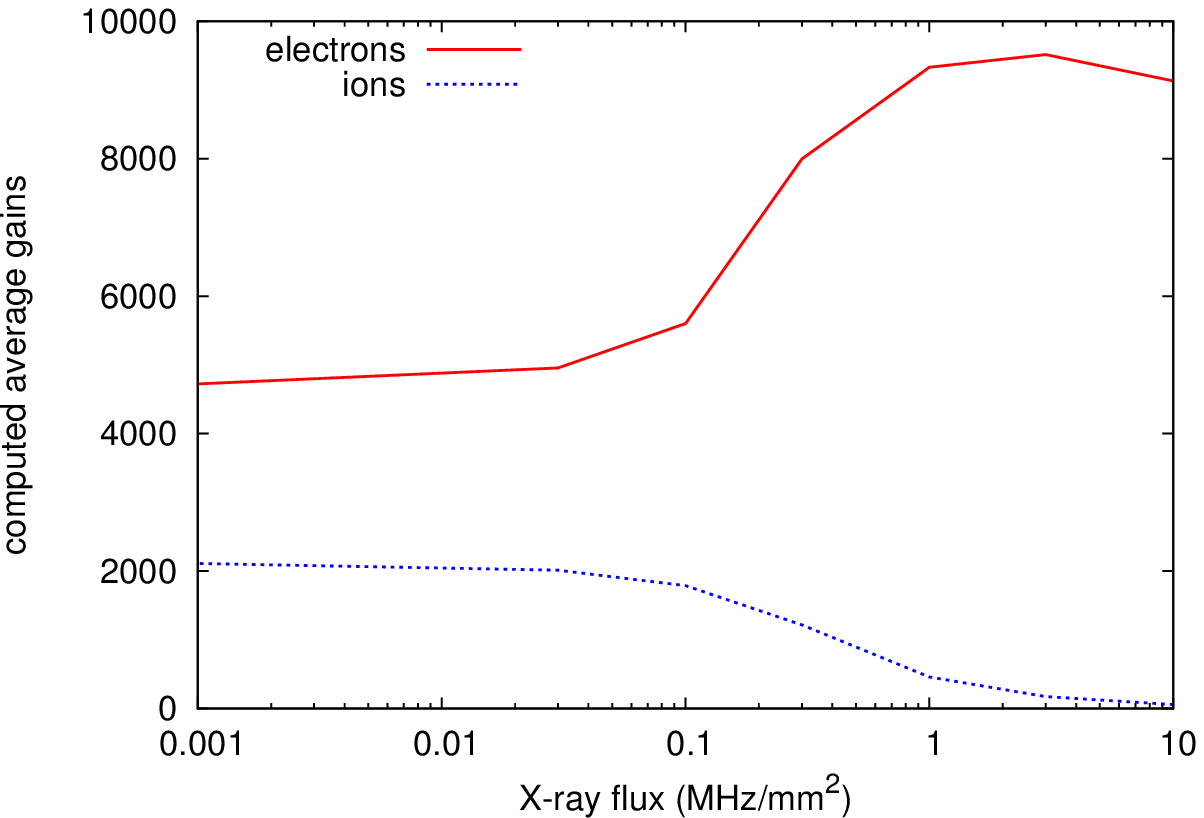}
\includegraphics[width=3.5in]{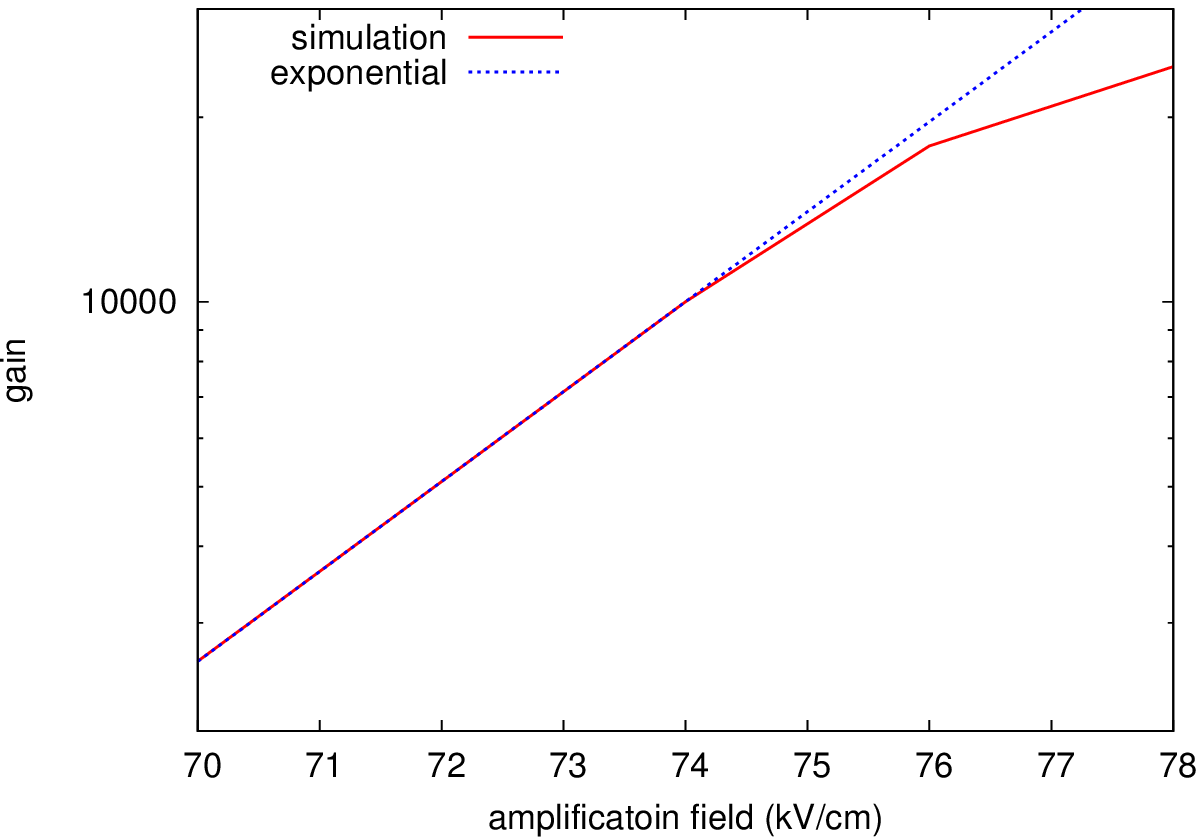}
\caption{Top: computed dependence of the average effective gain as a function of the average interaction X-ray flux, to be compared to figure~\ref{fig:gainVsFlux}. Bottom: calculated dependance of the effective gain as a function of the amplification field in each GEM, to be compared to figure~\ref{fig:gainVsVoltage} (bottom).}
\label{fig:calculated}
\end{figure}
The computed effective gain as a function of the X-ray flux is shown in figure~\ref{fig:calculated} (top) that must be compared to figure~\ref{fig:gainVsFlux}.
As can be noticed, the computation describes well the gain rise, but in the data the fall is more pronounced.
This can be due to neglected phenomena like the electron-ion recombination at high ion densities.
Nevertheless, we believe that the main source of discrepancy is the stochastic nature of the X-ray detection.
In the computation the primaries are generated continuously in time.
This can approximate an equally time spaced series of events.
In reality the time difference between an event and the previous is exponentially distributed.
Since the amplification drop is not a linear phenomenon with the electric field variations, it affects much more the events preceded close in time (and in space) by another event.

Concerning the gain saturation on an event by event basis, the same computation framework was used to calculate the effective gain of a single event, i.e.\ where the primaries are normally distributed in the middle of the conversion volume with an RMS of 100~$\mu$m.
The computed effective gain obtained by varying the nominal amplification field in the GEMs is shown in figure~\ref{fig:calculated} (bottom).
The deviation from the exponential curve reflects the behaviour observed in figure~\ref{fig:gainVsVoltage} (bottom).
The ions generated at the beginning of the avalanche in the last GEM move towards the \emph{entrance} of the hole, reducing the amplification field
and affecting the multiplication of the forthcoming electrons arriving from the transfer region.
Analogously to the effect described
previously, but with the difference that each
primary event now affects its own growth. 
The larger the avalanche is the more pronounced this effect is.
The discrepancy between the computed and the actual effective gain at which the saturation takes place must be attributed to the axisymmetric approximation of the geometry.
Multiple GEM reach much higher gain than a single GEM because between one stage and the other the charge is spread over several holes that behave as independent amplifiers.
What matters is therefore the amount of charge in the hole or, equivalently, the charge density.
In the computation the diffusion is taken into account, but only a single hole is described.
The approximation assumes that the charge exiting from the considered volume is compensated by the same amount of charge entering from the neighbouring volume.
This is valid when several holes are uniformly affected by the same amount of charge, as it happens for the high flux measurements.
It is certainly not valid for a single avalanche.
The saturation appears at lower gains in the computation because the charge per hole is overestimated.
\begin{figure}[!t]
\centering
\includegraphics[width=3.5in]{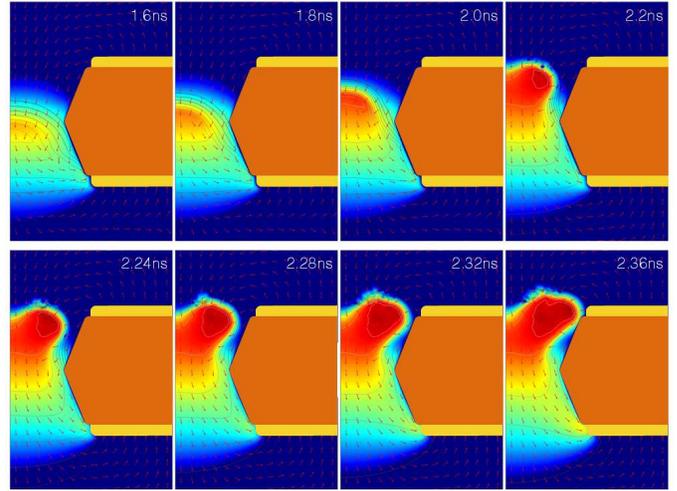}
\caption{Formation and propagation of a streamer in a GEM hole in cylindrical coordinates. The colour map represents the ion density in arbitrary units.}
\label{fig:discharge}
\end{figure}

Increasing further the gain beyond the onset of the saturation, a cathode-directed ionisation wave appears in simulation, something that we will term here as \emph{streamer}.
The remainder of this section is necessarily qualitative on this important aspect, that deserves further investigation.
Figure~\ref{fig:discharge} shows the formation and propagation of a streamer in a GEM hole computed within the same framework.
Thus also in the simulation there is a maximum achievable gain, after which the amount of electron-ion pairs diverges.
The figure shows the time sequence of the streamer happening in the hole of the GEM, displayed in cylindrical coordinates.
The colour map represents the ion density and the arrows show the direction of the electron drift.
The electric field in the hole is instantaneously and locally modified by the high negative and positive charge densities.
As described in~\cite{Fonte:2010,Resnati:2012}, the electric field increases in front of the ion cloud, upstream of the avalanche region before all the primary electrons have been fully amplified.
New incoming electrons generate in this region more electron-ion pairs that contribute to exacerbate the field modification.
In addition, a fraction of electrons present in the highest ionised region diffuses against the drift, populating the high field region in front of the ion cloud, and exacerbating even more the field distortion.
As a function of time the field keeps increasing towards the cathode and so does the ionisation rate.
This effectively results in a cathode-directed ionisation wave, i.e.\ a diffusion-assisted streamer~\cite{Ebert:1997}.
The streamer propagates \emph{backwards}, despite the electrons on average move towards the anode.
Following conventional wisdom, once the streamer reaches the top electrode of the GEM, a highly conducting channel between the two faces of the GEM is established in the gas, and the breakdown takes place.

\section{Conclusion}
A systematic series of measurements was performed to address the behaviour of GEM detectors in extreme particle fluxes and gain conditions.
The numerical model that we developed to validate the interpretation describes well the trends observed in our measurements.
The effect of positive charges in the transfer and amplification processes explains the observed behaviours.

Concerning the high flux behaviour, the intrinsic limits of triple GEM can be pushed forward by reducing the total space charge, for instance reducing the length of the transfer regions and using gases with larger ion mobility, e.g.\ neon based mixtures.
It is also possible to limit the influence of space charge, for instance by using GEMs with high electron transparency, e.g.\ larger hole to pitch ratio.
The increase of the effective gain is expected to be common to all the devices where the charge is not fully transferred from one stage to the next one.

The ions are responsible also for the observed saturation at high gains, but in this case the positive charge of each event quenches its amplification. 
This behaviour precedes in simulation the onset of a cathode-directed ionisation wave, that happens at yet higher gains and that can also be computed within the same numerical model.
We interpret this process as a cathode-directed streamer.

A more systematic set of data and a more detailed description of the modeling procedure will be included in a future publication, that is currently under preparation.

\section*{Acknowledgment}
The authors would like to acknowledge Francisco Garcia Fuentes and Ella Warras for the support on the initial measurements, Nayana Majumdar and Supratik Mukhopadhyay for encouraging measurements at high gains, and \"{O}zkan \c{S}ahin for providing the Penning-corrected values of the Townsend coefficient.

\end{document}